\documentclass[aps,pra,twocolumn,showkeys,amssymb,floatfix,longbibliography,superscriptaddress]{revtex4-1}

\usepackage{graphicx}
\usepackage{amsmath,amsfonts,amssymb}
\usepackage{hyperref}
\usepackage{braket}
\usepackage{xcolor}
\usepackage{tikz}
\usepackage{color}
\usepackage{enumerate}
\usepackage[caption=false]{subfig}
\usepackage{multirow}
\usepackage{qcircuit}
\usepackage[normalem]{ulem}
\usepackage{placeins}
\usepackage{url}

\DeclareRobustCommand\id{\leavevmode\hbox{\small1\normalsize\kern-.33em1}}
\newcommand{\bb}[1]{\ensuremath{\mathbb{#1}}}
\newcommand{\spn}[1]{\ensuremath{\func{span}\!\left\{#1\right\}}}
\newcommand{\f}[1]{\ensuremath{\mathrm{#1}\,}}

\newcommand{\func}[1]{\ensuremath{\mathrm{#1}}}

\newcommand{\appref}[1]{Appendix~\ref{#1}}
\newcommand{\ssecref}[1]{Section~\ref{#1}}
\newcommand{\secref}[1]{Sec.~\ref{#1}}
\newcommand{\figref}[1]{Fig.~\ref{#1}}
\newcommand{\tabref}[1]{Table~\ref{#1}}
\newcommand{\equref}[1]{Eq.~\eqref{#1}}
\newcommand{\equaref}[2]{Eqs.~\eqref{#1} and \eqref{#2}}
\newcommand{\equsref}[2]{Eqs.~\eqref{#1}--\eqref{#2}}

\newcommand{\mc}[1]{\ensuremath{\mathcal{#1}}}

\newcommand{\ds}{\,\mathrm{d}}
\newcommand{\da}{\mathrm{d}}
\newcommand{\p}{\partial}

\newcommand{\ketbra}[2]{\ket{#1}\!\!\bra{#2}}

\let\baraccent=\=
\renewcommand{\=}[1]{\stackrel{#1}{=}}

\newcommand\T{\rule{0pt}{2.6ex}}

\newlength\stextwidth
\newcommand\makesamesize[3][c]{\settowidth{\stextwidth}{\ensuremath{#2}}\makebox[\stextwidth][#1]{\ensuremath{#3}}}

\newcommand\SI[2]{{#1}\,\mathrm{#2}}

\begin{document}

\title{Gate-error analysis in simulations of quantum computers with transmon qubits}

\author{D. Willsch}
\affiliation{Institute for Advanced Simulation, J\"ulich Supercomputing Centre,\\
Forschungszentrum J\"ulich, D-52425 J\"ulich, Germany}
\author{M. Nocon}
\affiliation{Institute for Advanced Simulation, J\"ulich Supercomputing Centre,\\
Forschungszentrum J\"ulich, D-52425 J\"ulich, Germany}
\author{F. Jin}
\affiliation{Institute for Advanced Simulation, J\"ulich Supercomputing Centre,\\
Forschungszentrum J\"ulich, D-52425 J\"ulich, Germany}
\author{H. De Raedt}
\affiliation{Zernike Institute for Advanced Materials,\\
University of Groningen, Nijenborgh 4, NL-9747 AG Groningen, The Netherlands}
\author{K. Michielsen}
\affiliation{Institute for Advanced Simulation, J\"ulich Supercomputing Centre,\\
Forschungszentrum J\"ulich, D-52425 J\"ulich, Germany}
\affiliation{RWTH Aachen University, D-52056 Aachen, Germany}

\date{\today}

\begin{abstract}
  In the model of gate-based quantum computation, the qubits are controlled by a sequence of quantum gates. In superconducting
qubit systems, these gates can be implemented by voltage pulses. The success of implementing a particular gate can be expressed
by various metrics such as the average gate fidelity, the diamond distance, and the unitarity. We analyze these metrics of gate
pulses for a system of two superconducting transmon qubits coupled by a resonator, a system inspired by the architecture of the
IBM Quantum Experience. The metrics are obtained by numerical solution of the time-dependent Schr\"odinger equation of the
transmon system. We find that the metrics reflect systematic errors that are most pronounced for echoed cross-resonance gates,
but that none of the studied metrics can reliably predict the performance of a gate when used repeatedly in a quantum algorithm.
\end{abstract}

\keywords{quantum computation; superconducting qubits; quantum circuits; quantum error correction; quantum information; average fidelity; fault-tolerance thresholds; transmon qubits}

\maketitle

\section{Introduction}\label{sec:introduction}
Over the past decades, tremendous effort has gone into building a universal quantum computer. In theory, such a device can solve certain problems, such as factoring, exponentially faster than classical digital computers. The leading technological prototypes are based on superconducting transmon qubits containing on the order of 10 qubits \cite{gambetta2015building,kelly2015statepreservation9qubits,Song201710China10Qubits,ReagorRigetti2017universalparametric}. IBM provides public access to such a quantum processor through the IBM Quantum Experience (IBMQX) \cite{ibmquantumexperience2016}.

However, as reported in a recent independent benchmark \cite{Michielsen2017BenchmarkingQC}, IBM's five-qubit quantum processor does not yet meet the fundamental requirements for a computing device. For this reason, the underlying architecture and its operation call for a deeper analysis, one that goes beyond perturbation theory, rotating wave approximations, and assumptions about Lindblad forms and Markovian dynamics \cite{Puzzuoli2014tractablesimulation}.

We study the real-time dynamics of such quantum systems in detail by solving the time-dependent Schr\"odinger equation (TDSE) for a generic model Hamiltonian. For this purpose, we have developed efficient product-formula algorithms that are tailored to key features of the model Hamiltonian \cite{deraedt1987productformula}. This allows us to simulate each Gaussian control pulse that is used in experiments to realize a certain quantum gate, as dictated by the computational model of a quantum computer. We have implemented a parameter optimization scheme for obtaining the best pulse parameters for the gates on the transmon system. This scheme makes use of the fact that in the simulation, we have the advantage of getting the full information of the system dynamics at any time $t$. In brief, the simulated system can be seen as a faithful model of an ideal quantum processor that works exactly as quantum theory dictates.

In this paper, we limit the analysis to two transmons coupled by one resonator as fundamental errors can be best understood for a small system containing only the basic constituents. For the implemented gates, we evaluate the average gate fidelity \cite{nielsen2002gatefidelity}, the diamond distance \cite{kitaev1997diamondnorm}, and the unitarity \cite{Wallman2015unitarity}. The obtained gate fidelities agree with those reported in state-of-the-art experiments \cite{sheldon2015singlequbitfidelities,sheldon2016procedure,barends2014superconducting,kelly2014optimalcontrolrb}. However, we find that the diamond error rates of all gates are larger than 2\% (see also \cite{Puzzuoli2014tractablesimulation}). The precision of the gates is limited by the presence of non-computational states in the transmons and the resonator. The corresponding errors occur naturally in the unitary evolution of the total system, but they have a detrimental effect on the computational subspace. For instance, we find that they appear incoherent when looking at the computational subspace only, and they cannot be represented by Pauli channels \cite{NielsenChuang}.

In particular, for controlled-NOT (CNOT) gates based on echoed cross-resonance pulses \cite{chow2011CR,rigetti2010CR,Corcoles2013processverification}, we find a systematic error that can be reproduced in experiments on the IBMQX. We also propose a different, one-pulse CNOT gate that does not suffer from this error.

The paper is structured as follows. In \secref{sec:model}, we describe the simulation model and explain how quantum gates are implemented by microwave pulses. This section also sketches the optimization procedure that we use to find optimal pulses for the qubit system. \ssecref{sec:errorrates} gives a summary of the gate metrics that frequently serve as error rates in experimental and theoretical studies. In \secref{sec:results}, we present the gate metrics of the optimized pulses and analyze their behavior with respect to repeated applications of the gates. Additionally, we perform identity operations and entanglement experiments as proposed in \cite{Michielsen2017BenchmarkingQC} to compare the performance of the gate sets with ``real'' quantum programs. Conclusions drawn from our analysis are given in \secref{sec:discussion}.

\section{Simulation model and method}\label{sec:model}
We consider a system of superconducting transmon qubits \cite{koch2007transmon}. The transmons are coupled by a transmission line resonator, which is essentially a quantum harmonic oscillator \cite{blais2004circuitqed}. The publicly accessible five-qubit quantum processor of the IBMQX is of this type \cite{ibmquantumexperience2016}.

The model Hamiltonian of a system of $N$ transmons coupled to a resonator reads \cite{koch2007transmon}
\begin{align}
  \label{eq:modelhamiltonian}
  H &= H_{\text{CPB}} + H_{\text{res}},\\
  \label{eq:modelhamiltonianCPB}
  H_{\text{CPB}} &= \sum\limits_{i=1}^{N} \left[ E_{Ci}(\hat n_i - n_{gi}(t))^2 - E_{Ji} \cos\hat\varphi_i \right],\\
  \label{eq:modelhamiltonianRes}
  H_{\text{res}} &= \omega_r\hat a^\dagger\hat a + \sum\limits_{i=1}^{N} g_i \hat n_i(\hat a + \hat a^\dagger).
\end{align}
Here, $H_{\text{CPB}}$ describes the Cooper pair box (CPB) qubits whose capacitive energies $E_{Ci}$ and Josephson energies $E_{Ji}$ are set in the transmon regime \cite{koch2007transmon}, $\hat{n}_i$ is the number operator of qubit $i$, and the bounded phase operator $\hat\varphi_i$ is its conjugate. The qubits are controlled by the external control field $n_{gi}(t)$, which is directly proportional to the voltage pulses applied to the qubit. Thus, quantum gates are implemented by choosing a certain pulse form for $n_{gi}(t)$ (see \cite{gambetta2013controlIFF,McKay2016VZgate}). The resonator is described by raising and lowering operators $\hat a^\dagger$ and $\hat a$, respectively. It operates at the microwave frequency $\omega_r$ and its capacitive coupling strength to the qubits is given by $g_i$.

The values of the parameters in \equsref{eq:modelhamiltonian}{eq:modelhamiltonianRes} are given in \tabref{tab:device}. In what follows, we consider the case $N=2$ as the key results are most clearly demonstrated for a small isolated system of qubits.
\begin{table}
  \caption{\label{tab:device}Parameters for the model Hamiltonian given in \equsref{eq:modelhamiltonian}{eq:modelhamiltonianRes}, inspired by the device parameters of the quantum processor of the IBMQX \cite{ibmquantumexperience2016}. All energies are expressed in GHz ($\hbar=1$). The CPB qubits are operated in the transmon regime with $E_{Ji}/E_{Ci}\approx10$, and their frequencies $\omega_i$ and anharmonicities $\alpha_i$ resulting from diagonalizing the CPB Hamiltonian are given for reference. The resonator operates at frequency $\omega_r/2\pi=\SI{7}{GHz}$. Its coupling to the qubits is weak as $|g_i|\ll|\omega_i-\omega_r|$.}
\begin{ruledtabular}
\begin{tabular}{@{}cccccc@{}}
  Qubit $i$ & $E_{Ci}/2\pi$ & $E_{Ji}/2\pi$ & $\omega_i/2\pi$ & $\alpha_i/2\pi$ & $g_i/2\pi$\\
  \colrule\T
  1 & 1.204 & 13.349 & 5.350 & $-$0.350 & 0.07 \\
  2 & 1.204 & 12.292 & 5.120 & $-$0.353 & 0.07 \\
\end{tabular}
\end{ruledtabular}
\end{table}

The dynamics of the joint system of the two transmons and the resonator can be obtained by studying the time evolution of the state $\ket{\Psi(t)}$ of the system. This state is the solution of the TDSE ($\hbar=1$)
\begin{align}
  i\frac{\p}{\p t}\ket{\Psi(t)} &= H(t) \ket{\Psi(t)},
  \label{eq:TDSE}
\end{align}
for the Hamiltonian given in \equref{eq:modelhamiltonian}. We obtain the solution numerically by implementing a product-formula algorithm for the total unitary time-evolution operator $U_{\text{total}}(t)$ defined by $\ket{\Psi(t)} = U_{\text{total}}(t)\ket{\Psi(0)}$ (see \appref{app:algorithm} for details on the algorithm). This solution is expanded in the product basis $\ket{k}\!\ket{m_1}\!\ket{m_2}$ where $k$ is the number of photons in the resonator, and $m_i=0,1,2,\ldots$ label the transmon eigenstates (i.e., the eigenstates of $H_{\text{CPB}}$ given by \equref{eq:modelhamiltonianCPB} for $n_{gi}(t)=0$) of qubit $i=1,2$. Thus, the result of the simulation is the set of coefficients $a_{km_1m_2}(t)$ defined by
\begin{align}
  \ket{\Psi(t)} = \sum\limits_{km_1m_2}^{}a_{km_1m_2}(t) \ket{k}\!\ket{m_1}\!\ket{m_2}.
  \label{eq:TDSEsolution}
\end{align}
The system is initialized in a computational basis state $\ket{\Psi(0)}=\ket{m_1m_2}$, where the computational subspace is defined by $\ket{m_1m_2}=\ket{k=0}\!\ket{m_1}\!\ket{m_2}$ for $m_1,m_2\in\{0,1\}$. Note that the simulation explicitly includes non-computational states outside of this subspace.

\subsection{Quantum gates}
For architectures of the transmon type, quantum gates are implemented by applying microwave voltage pulses to the qubits. This is mathematically modeled through the control fields $n_{gi}(t)$ in \equref{eq:modelhamiltonianCPB}. We consider a generic sum of shaped microwave pulses applied on each qubit as described in \cite{McKay2016VZgate}, namely
\begin{align}
  n_{gi}(t) &= \sum\limits_{j}^{} \Omega_{ij}(t) \cos( \omega^{\mathrm{dr}}_{ij} t - \gamma_{ij} ),
  \label{eq:genericpulses}
\end{align}
where $\Omega_{ij}(t)$ is the envelope of pulse $j$ on qubit $i$, $\omega^{\mathrm{dr}}_{ij}$ is the corresponding drive frequency, and $\gamma_{ij}$ is an offset phase. To model a situation close to experiments (cf. \cite{motzoi2009drag,gambetta2013controlIFF}), the envelopes $\Omega_{ij}(t)$ are Gaussians of the form
\begin{align}
  \Omega_{\text{G}}(t) &= \Omega_0\frac{\exp\Big(-\frac{(t-T/2)^2}{2\sigma^2}\Big) - \exp\Big(-\frac{T^2}{8\sigma^2}\Big)}{1 - \exp\Big(-\frac{T^2}{8\sigma^2}\Big)},
  \label{eq:gauss}
\end{align}
where $\Omega_0$ is the amplitude and $T$ the time of the pulse, and $\sigma=T/4$ defines the width of the Gaussian.

The drive frequencies $\omega_{ij}^{\mathrm{dr}}$ in \equref{eq:genericpulses} are usually set to one of the qubit frequencies $\omega_i$ (see \tabref{tab:device}). However, as the presence of the resonator can slightly shift these frequencies \cite{blais2004circuitqed}, we adjust $\omega_i$ by measuring local rotations of the qubits in the laboratory frame. We do this by initializing the system in the state $\ket{\Psi(0)}=\ket{++}$ with $\ket{+}=(\ket{0}+\ket{1})/\sqrt{2}$ and letting it evolve freely for $\SI{4000}{ns}$. On the respective Bloch spheres, both qubits then rotate about the $z$ axis. The frequencies of these rotations yield the shifted qubit frequencies $\bar\omega_i$. We obtain $\bar\omega_1/2\pi=\SI{5.346}{GHz}$ and $\bar\omega_2/2\pi=\SI{5.118}{GHz}$.

The computational states of the qubits at some time $t>0$ are defined in a so-called \emph{locally rotating frame} $R$ \cite{NielsenChuang,gambetta2013controlIFF}. This essentially removes the just mentioned rotation, so that the state $\ket{++}$ remains unchanged if no quantum gate is applied. Mathematically, this is implemented by multiplying the coefficients of the solution given in \equref{eq:TDSEsolution} by time-dependent phase factors, yielding $a_{km_1m_2}^R(t) := \exp(it(\bar\omega_1m_1+\bar\omega_2m_2)) a_{km_1m_2}(t)$.

We have implemented the same quantum gate set as supported by the IBMQX \cite{Cross2017openqasm2}. Accordingly, a typical quantum gate sequence takes between $\SI{50}{ns}$ and $\SI{15}{\mu s}$. In the following, we explain how the pulses are defined and modeled.

\subsubsection{Single-qubit gates}
Single-qubit rotations on the Bloch sphere can be realized by applying a Gaussian pulse with drive frequency $\omega_i^{\mathrm{dr}}=\bar\omega_i$ on qubit $i$ (see \equaref{eq:genericpulses}{eq:gauss}). In this case, the amplitude $\Omega_0$ and the phase $\gamma$ define the angle and the axis of rotation, respectively \cite{gambetta2013controlIFF} (e.g., $\gamma=0$ for rotations about the $x$ axis, or $\gamma=\pi/2$ for rotations about the $y$ axis).

We utilize the virtual Z gate (VZ gate) described in \cite{McKay2016VZgate} and used in the IBMQX \cite{ibmquantumexperience2016} to implement rotations about the $z$ axis. This means that instead of applying another pulse, we simply change the phase $\gamma$ of all the following pulses (see \appref{app:gatepulsedetails} for details).

Unfortunately, as transmons cannot be represented by ideal two-level systems, such pulses may induce leakage out of the computational subspace, meaning that the solution in \equref{eq:TDSEsolution} also has contributions from higher levels such as $\ket{m_i=2}$. This effect can be mitigated by including another pulse in \equref{eq:genericpulses} proportional to the derivative $\dot\Omega_{\text{G}}(t)$ with a phase shift of $\pi/2$. This technique goes under the name of DRAG and has become standard for transmon systems \cite{motzoi2009drag,gambetta2010dragtheory}. Therefore we also adopt DRAG in defining the pulses.

For the single-qubit gates, we take $T=\SI{83}{ns}$ for the gate duration of the Gaussian envelope $\Omega_G(t)$ given by \equref{eq:gauss}, inspired by the choice made for the IBMQX \cite{ibmquantumexperience2016}.

In summary, a single-qubit pulse on qubit $i$ is defined by
\begin{align}
  n_{gi}(t) &= \Omega_{\text{G}}(t)\cos(\bar\omega_it-\gamma) + \beta\dot\Omega_{\text{G}}(t)\cos(\bar\omega_it-\gamma-\frac{\pi}{2}),
  \label{eq:singlequbitpulse}
\end{align}
with the parameters $(\Omega_0,\beta,\gamma)$ being the amplitude, the DRAG coefficient, and the phase, respectively (see \appref{app:gatepulsedetails} for the theory behind these parameters). As outlined below, we optimize the pulse parameters to implement ideal single-qubit rotations of the type $X_{\pi/2}$ and $X_{\pi}$. The former serves as a primitive to generate arbitrary single-qubit gates as in experiments \cite{McKay2016VZgate}. The latter is used exclusively as a component in the echoed two-qubit gates.

\subsubsection{Two-qubit gates}
We implement the CNOT gate by making use of the cross-resonance (CR) effect \cite{chow2011CR,rigetti2010CR}. The basic idea simply amounts to applying another Gaussian pulse to the control qubit $C=1,2$ but at the drive frequency $\bar\omega_T$ of the target qubit $T\neq C$. Furthermore, the pulse is stretched over a longer time period such that the Gaussian in \equref{eq:gauss} becomes a flat-topped Gaussian with $3\sigma$ rise time where $\sigma=\SI{5}{ns}$ (cf. \cite{sheldon2016procedure}).

\begin{figure}
  \includegraphics[width=\linewidth]{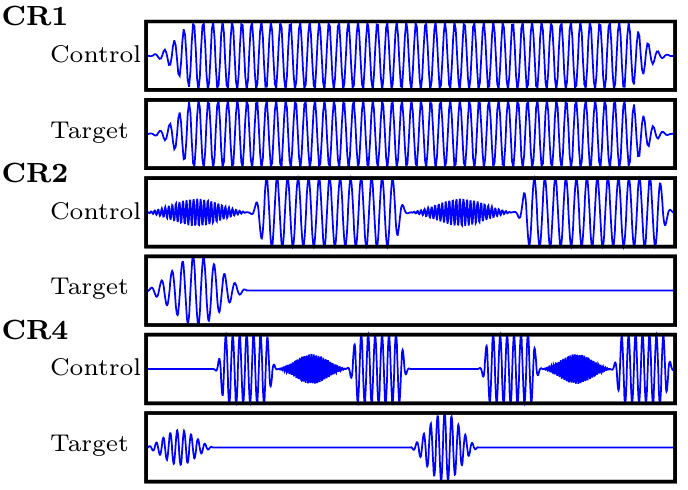}
  \caption{\label{fig:CNOTpulses}Pulse sequences for the three different realizations of a CNOT gate studied in this paper. Gaussians implement $X_{\pi/2}$ and $X_{\pi}$ rotations, and flat-topped Gaussians represent cross-resonance (CR) pulses (i.e., they oscillate at the frequency $\bar\omega_T$ of the target qubit). The CR1 gate consists only of flat-topped Gaussian pulses at the target frequency $\bar\omega_T$. The CR2 gate is an echoed CR gate containing two additional $X_{\pi}$ pulses on the control qubit and one $X_{\pi/2}$ pulse on the target qubit. The CR4 gate is a four-pulse echoed CR gate that contains an additional $X_{\pi}$ pulse on the target qubit. See \figref{fig:twoqubitcircuits} in \appref{app:gatepulsedetails} for the full pulse specifications.}
\end{figure}

Various schemes have been used in experiments to construct a CNOT gate based on the CR effect \cite{Corcoles2013processverification,corcoles2015demonstration,takita2016demonstration,Takita2017faultTolerantStatePreparation,sheldon2016procedure}. We implement three particularly interesting representatives to compare their performance with the performance of the ideal system. The first is a simple one-pulse CR gate (CR1) that we found by including an additional driving of the target qubit, inspired by the observation in \cite{sheldon2016procedure} (see \figref{fig:CRamplitudescan} in \appref{app:gatepulsedetails} for details). The second is a two-pulse echoed CR gate (CR2) which is currently also used on the five-qubit quantum processor of the IBMQX \cite{ibmquantumexperience2016}. The third is a four-pulse echoed CR gate (CR4) that has recently shown better performance (albeit worse fidelity) for an experiment on quantum error-detecting codes \cite{Takita2017faultTolerantStatePreparation}. The pulse sequences of the three CNOT gates are summarized in \figref{fig:CNOTpulses}.

The pulse parameters such as amplitudes, times, and phases for each sequence are scanned over ranges suggested by the theory. As in the case of single-qubit gates, this provides initial values for the pulse optimization procedure.

\subsection{Pulse optimization}\label{sec:pulseoptimization}
The goal of applying the control pulses is to realize a certain transformation $U$ (the unitary quantum gate) on the computational subspace. For two qubits, this subspace is spanned by the computational basis $\{\ket{00},\ket{01},\ket{10},\ket{11}\}$, so $U$ is essentially a complex $4\times4$ matrix. Examples for $U$ are $X^i_{\pi/2}$ (a $\pi/2$ rotation of qubit $i$ about the $x$ axis) or $\mathrm{CNOT}_{ij}$ (a controlled NOT operation where $i,j$ with $i\neq j$ denote the control and the target qubit, respectively) \cite{NielsenChuang}.

As the simulation produces the full state $\ket{\Psi(t)}$ given by \equref{eq:TDSEsolution}, we can construct the actual transformation matrix $M$ implemented by a certain pulse. We do this by initializing $\ket{\Psi(0)}$ in each of the four computational basis states, evolving the system under the application of the pulse according to \equref{eq:TDSE}, and extracting the four complex coefficients $a_{000}$ through $a_{011}$ from the solution given by \equref{eq:TDSEsolution} (formally, $M$ is a $4\times4$ block of the total time-evolution operator $ U_{\text{total}}(t)$, see \equref{eq:fulltimeevolutionoperator}). Each run for one of the four computational basis states produces one column of $M$, including the complex phases that each basis state acquires in the time evolution. The four runs can be performed in parallel.

The aim is to find ideal pulse parameters so that the computational matrix $M$ approaches the ideal gate matrix $U$, up to a global phase. Note that the computational space is a subspace of the whole Hilbert space $\mc H=\spn{\ket{k}\!\ket{m_1}\!\ket{m_2}}$, so it is by no means clear that $M$ will be unitary (and in almost all cases, it is not).

We use a multidimensional optimization scheme introduced by Nelder and Mead \cite{NelderMead1965,numericalrecipes} to optimize the pulse parameters. Note that we only use the optimization procedure to refine the initial pulse parameters obtained from the theory \cite{gambetta2010dragtheory,McKay2016VZgate,sheldon2016procedure} (see also \appref{app:gatepulsedetails}). The objective function to be minimized is given by
\begin{align}
  \Delta(M,U) &= \left\|M - z U\right\|_F^2,
  \label{eq:distance}
\end{align}
where $\left\|\cdot\right\|_F$ is the Frobenius norm, and $z=\pm\sqrt{\func{Tr}(MU^\dagger)/\func{Tr}(MU^\dagger)^*}$ is a phase factor that minimizes the global phase difference between both matrices. We have tested other gate error rates as objective functions (see \secref{sec:errorrates}) and found that \equref{eq:distance} produces the best results.

After optimizing the pulse parameters, we further improve the gates using the VZ phase correction to compensate for off-resonant rotation errors, etc. \cite{McKay2016VZgate} (see \appref{app:gatepulsedetails} for details).

\section{Gate error rates}\label{sec:errorrates}
Various quantities have been used in experiments and studied in the literature to measure the success of implementing a quantum gate by a certain control pulse \cite{Wallman2015ErrorRates}. Some of these measures are motivated by their simplicity and straightforwardness in the experimental implementation (e.g., average gate fidelity \cite{nielsen2002gatefidelity}), while others such as the diamond distance stem from mathematical considerations \cite{kitaev1997diamondnorm}. As recently demonstrated by Sanders \emph{et~al.} \cite{Sanders2016ThresholdTheorem}, the relation between fidelity and diamond distance is not direct in that the impressively high fidelities reported in experiments are not sufficient for fault-tolerant quantum computation, in contrast to claims made by other groups \cite{chowGambetta2012fidelitiesandcoherence,barends2014superconducting}

In the following, we give an overview of the three metrics that we have selected to assess the quality of quantum gates. Evaluating these metrics requires the definition of appropriate \emph{quantum channels}, which are completely positive (CP) linear maps on the space of density operators $\rho$ \cite{breuer2007openquantumsystems}. For a two-qubit system, $\rho$ is a Hermitian $4\times4$ matrix. Using the language from \secref{sec:pulseoptimization} where $U$ denotes the ideal unitary gate matrix and $M$ denotes the actual transformation implemented on the computational subspace, we define the ideal quantum channel $\mc G_{id}$ and the actual quantum channel $\mc G_{ac}$ as
\begin{align}
  \label{eq:Gideal}
  \mc G_{id}(\rho) &= U\rho U^\dagger,\\
  \label{eq:Gactual}
  \mc G_{ac}(\rho) &= M\rho M^\dagger.
\end{align}
It can easily be seen that both maps are CP. However, note that in most cases $M^\dagger\neq M^{-1}$ because of additional non-computational states present in transmon systems. This implies that $\mc G_{ac}$ is not trace-preserving (the alternative channel $M\rho M^{-1}$ does not preserve Hermiticity).

For convenience, we define the discrepancy channel $\mc D(\rho) = \mc G_{ac}(\mc G_{id}^{-1}(\rho))$ which approaches unity for a perfect control pulse.

\subsection{Average gate fidelity}
The average gate fidelity is defined as \cite{nielsen2002gatefidelity}
\begin{align}
  \label{eq:fidelity}
  F_{\text{avg}} &= \int \mathrm{d}\!\ket{\psi}\, \bra{\psi} \mc D(\ketbra\psi\psi) \ket{\psi}.
\end{align}
In general, we have $0\le F_{\text{avg}}\le1$, and the maximum fidelity $F_{\text{avg}}=1$ is attained in the ideal case where the discrepancy channel $\mc D$ is unity.

In experiments, this number is estimated by a protocol called randomized benchmarking (RB) \cite{emerson2005randomunitaryoperators,Magesan2012RBandDiamondNorm}. However, as in our simulation we have access to the error channel given by \equref{eq:Gactual}, we do not need to implement the RB protocol. Instead, we evaluate \equref{eq:fidelity} directly by sampling the integrand and averaging it over states from the computational subspace, as done in \cite{Wood2017LeakageRB}. Specifically, we generate 100,000 random states $\ket{\psi}=\sum_{ij}^{}c_{ij}\ket{ij}$ by drawing real and imaginary parts of $c_{ij}$ from a normal distribution and normalizing the state afterwards.

\subsection{Diamond distance}
The error rate of a quantum operation is defined in terms of the diamond distance \cite{Sanders2016ThresholdTheorem}
\begin{align}
  \eta_\Diamond &= \frac{1}{2} \left\|\mc D - \id\right\|_\Diamond.
  \label{eq:errorrate}
\end{align}
This quantity is mathematically relevant for the \emph{threshold theorem} \cite{aharonov2008thresholdtheorem} that is often cited in the literature to argue that arbitrarily long, fault-tolerant quantum computation is possible. The best known quantum error-correcting codes require $\eta_\Diamond$ to be on the order of 1\% or less \cite{Sanders2016ThresholdTheorem}.

Evaluating \equref{eq:errorrate} is nontrivial as the diamond norm of a superoperator $\mc A$ is defined by maximizing the trace norm $\left\|\cdot\right\|_{\func{Tr}}$ over all ancillary Hilbert spaces $\mc H'$ and all joint density operators on $\mc H\otimes\mc H'$ \cite{kitaev1997diamondnorm,Sanders2016ThresholdTheorem}. However, one can show that this is equivalent to minimizing over all generalized Choi-Kraus representations of $\mc A$ \cite{Johnston2009ComputingStabilizedNormsQC}. As we have $\mc A(\rho)=MU^\dagger\rho UM^\dagger-\rho$, this amounts to computing
\begin{align}
  \eta_\Diamond = \frac1 2 \inf_{S} \left\{ \left\| \begin{pmatrix}
    UM^\dagger, & -\id
  \end{pmatrix} S^{-\dagger} S^{-1} \begin{pmatrix}
    MU^\dagger \\
    -\id
  \end{pmatrix}\right\|_2^{1/2} \right.\nonumber\\
  \left. *\left\| \begin{pmatrix}
    UM^{\dagger}, & \id
  \end{pmatrix} S S^\dagger \begin{pmatrix}
    MU^\dagger \\
    \id
  \end{pmatrix}\right\|_2^{1/2} \right\}.
  \label{eq:errorratecomputation}
\end{align}
Here, $\left\|\cdot\right\|_2$ denotes the matrix 2-norm (i.e. the maximum singular value \cite{golub1996matrix}), and $S$ is an invertible complex $2\times2$ matrix. We solve this minimization problem by first sampling over 10,000 random matrices $S$ and then running the same optimization procedure that we already implemented for the pulse optimization in \secref{sec:pulseoptimization}. This scheme was found to produce reliable results, equal to the exact $\eta_\Diamond$ up to two significant digits for all tested cases for which we found closed expressions (see \cite{Johnston2009ComputingStabilizedNormsQC}).

There are two asymptotically tight bounds for the error rate $\eta_\Diamond$ in terms of the average gate fidelity $F_{\text{avg}}$ given by \equref{eq:fidelity} \cite{Sanders2016ThresholdTheorem}, namely
\begin{align}
  \label{eq:etapauli}
  \eta_\Diamond^{\text{Pauli}} &= \frac{d+1}{d}(1-F_{\text{avg}}),\\
  \label{eq:etaub}
  \eta_\Diamond^{\text{ub}} &= \sqrt{d(d+1)(1-F_{\text{avg}})},
\end{align}
for which we have $\eta_\Diamond^{\text{Pauli}}\le\eta_\Diamond\le\eta_\Diamond^{\text{ub}}$. In these expressions, $d=2^N=4$ is the dimension of the computational subspace. The upper bound leads to the estimate that two-qubit gates need to reach a fidelity of 0.999995 in order to qualify for fault-tolerant quantum computation with known quantum error-correcting codes \cite{Sanders2016ThresholdTheorem}. The lower bound is saturated if the error is a Pauli channel, and the difference $\eta_\Diamond-\eta_\Diamond^{\text{Pauli}}$ represents the ``badness'' of the noise. We shall see that all gates under investigation yield $\eta_\Diamond\gg\eta_\Diamond^{\text{Pauli}}$.

\subsection{Unitarity}
For transmon qubits, leakage into higher non-computational levels during the application of a pulse is a known problem \cite{chen2016leakagemartinis,Wood2017LeakageRB}. Mathematically, this leads to the situation that the evolution of the computational subspace is not unitary, resulting in $M^\dagger\neq M^{-1}$ and thus $\f{Tr} \mc G_{ac}(\rho) < \f{Tr}\rho$ in terms of \equref{eq:Gactual}, so the process is not trace-preserving. To quantify such effects, Wallman \emph{et~al.} have proposed a quantity called \emph{unitarity} \cite{Wallman2015unitarity} given by
\begin{align}
  \label{eq:unitarity}
  u &= \frac{d}{d-1} \int \mathrm{d}\!\ket{\psi}\, \func{Tr}\left[ \mc G_{ac}'(\ketbra\psi\psi)^\dagger \mc G_{ac}'(\ketbra\psi\psi) \right],
\end{align}
where $\mc G_{ac}'(\rho) = \mc G_{ac}(\rho-\id/d) - \f{Tr}\left[\mc G_{ac}(\rho-\id/d)\right]/\sqrt{d}$.

Note that by construction, the errors we observe are systematic, unitary, and coherent (in the sense of \cite{Wallman2015unitarity}) on the total Hilbert space $\mc H$. Hence this quantity characterizes how incoherent these errors appear on the computational subspace.

The integral in \equref{eq:unitarity} is evaluated in the same way as the average gate fidelity given in \equref{eq:fidelity}.

\section{Results}\label{sec:results}
In this section, we analyze the performance of the optimized single-qubit and two-qubit quantum gate sets. First, we evaluate the gate metrics described in the previous section. Then we study the repeated application of gates that mathematically constitute identity operations. Finally, we repeat a set of quantum entanglement experiments to compare the simulated results with the corresponding experimental results obtained on the IBMQX \cite{Michielsen2017BenchmarkingQC}.

\subsection{Gate metrics}
\begin{table*}
  \caption{\label{tab:metrics}Gate metrics for the set of optimized quantum gate pulses described in \secref{sec:model}. The distance objective $\Delta$ from the optimization is given in \equref{eq:distance}. The average gate fidelity $F_{\text{avg}}$, the error rates $\eta_\Diamond,\eta_\Diamond^{\text{Pauli}},\eta_\Diamond^{\text{ub}}$, and the unitarity $u$ are defined in Eqs.~\eqref{eq:fidelity}, \eqref{eq:errorrate}, \eqref{eq:etapauli}, \eqref{eq:etaub}, and \eqref{eq:unitarity}, respectively.}
\begin{ruledtabular}
\begin{tabular}{clccccccc@{}}
  Type & Gate & $T$ in ns & $\Delta$ & $F_{\text{avg}}$ & $\eta_\Diamond$ & $\eta_\Diamond^{\text{Pauli}}$ & $\eta_\Diamond^{\text{ub}}$ & $u$\\
  \colrule\T
  \multirow{4}{*}{\bfseries X}   & $X_{\pi/2}^1$        &                 83 &    0.0022 &      0.9946 &      0.027 &    0.0068 &      0.33 &      0.990 \\
                                 & $X_{\pi/2}^2$        &                 83 &    0.0023 &      0.9942 &      0.028 &    0.0073 &      0.34 &      0.989 \\
                                 & $X_{\pi}^1$          &                 83 &    0.0013 &      0.9949 &      0.020 &    0.0064 &      0.32 &      0.990 \\
                                 & $X_{\pi}^2$          &                 83 &    0.0015 &      0.9943 &      0.023 &    0.0071 &      0.34 &      0.989 \\
  \colrule\T
  \multirow{2}{*}{\bfseries CR1} & $\mathrm{CNOT}_{12}$ & \hphantom{0}71.865 &    0.0013 &      0.9842 &      0.029 &    0.0198 &      0.56 &      0.969 \\
                                 & $\mathrm{CNOT}_{21}$ &            158.193 &    0.0023 &      0.9951 &      0.033 &    0.0062 &      0.31 &      0.991 \\
  \colrule\T
  \multirow{2}{*}{\bfseries CR2} & $\mathrm{CNOT}_{12}$ &            431.949 &    0.0061 &      0.9943 &      0.048 &    0.0071 &      0.34 &      0.991 \\
                                 & $\mathrm{CNOT}_{21}$ &            369.116 &    0.0056 &      0.9947 &      0.048 &    0.0066 &      0.32 &      0.992 \\
  \colrule\T
  \multirow{2}{*}{\bfseries CR4} & $\mathrm{CNOT}_{12}$ &            652.954 &    0.0054 &      0.9934 &      0.049 &    0.0083 &      0.36 &      0.989 \\
                                 & $\mathrm{CNOT}_{21}$ &            572.623 &    0.0045 &      0.9946 &      0.044 &    0.0068 &      0.33 &      0.991 \\
\end{tabular}
\end{ruledtabular}
\end{table*}

The gate metrics of the optimized pulses are given in \tabref{tab:metrics}. The overall performance of the pulses is close to optimal but still not perfect. Especially the error rate $\eta_\Diamond$ given by \equref{eq:errorrate} is always bounded above $2\%$, even though our quantum-theoretical model of the transmon qubit architecture does not account for decoherence or noise. The average gate fidelities are in the same ballpark as those reported for experiments based on the same pulse schemes \cite{gambetta2015building,sheldon2015singlequbitfidelities,sheldon2016procedure,kelly2014optimalcontrolrb,barends2014superconducting}. In fact, the single-qubit gate fidelities are slightly worse than the ones reported in experiments. We shall see below, however, that the actual performance of the gates in quantum circuits is much better. We also observed similar gate metrics for shorter single-qubit gates of about $T=\SI{10}{ns}$, but then the performance of repeated applications of the pulses was worse (data not shown).

Note that we always find $\eta_\Diamond\gg\eta_\Diamond^{\text{Pauli}}$, so the dominant errors are non-Pauli errors and belong to the ``bad'' class of errors \cite{Sanders2016ThresholdTheorem}. Interestingly, there is almost one order of magnitude difference between the actual error rate $\eta_\Diamond$ and the optimal bounds $\eta_\Diamond^{\text{Pauli}}$ and $\eta_\Diamond^{\text{ub}}$ calculated from the gate fidelity $F_{\text{avg}}$ according to \equaref{eq:etapauli}{eq:etaub}.

In \tabref{tab:metrics}, it is also seen that a higher fidelity $F_{\text{avg}}$ corresponds to a higher unitarity $u$. From this we conclude that leakage is still the dominant source of error limiting the gate fidelity, even though the techniques DRAG \cite{motzoi2009drag} and VZ phase correction \cite{McKay2016VZgate} have been included in the construction of the pulses. It seems that the presence of the resonator and the entangling transverse coupling cause this limitation (see also \cite{billangeon2015longitudinal,richer16longitudinal}).

\subsection{Repeated gate applications}
\begin{figure}
  \includegraphics[width=\linewidth]{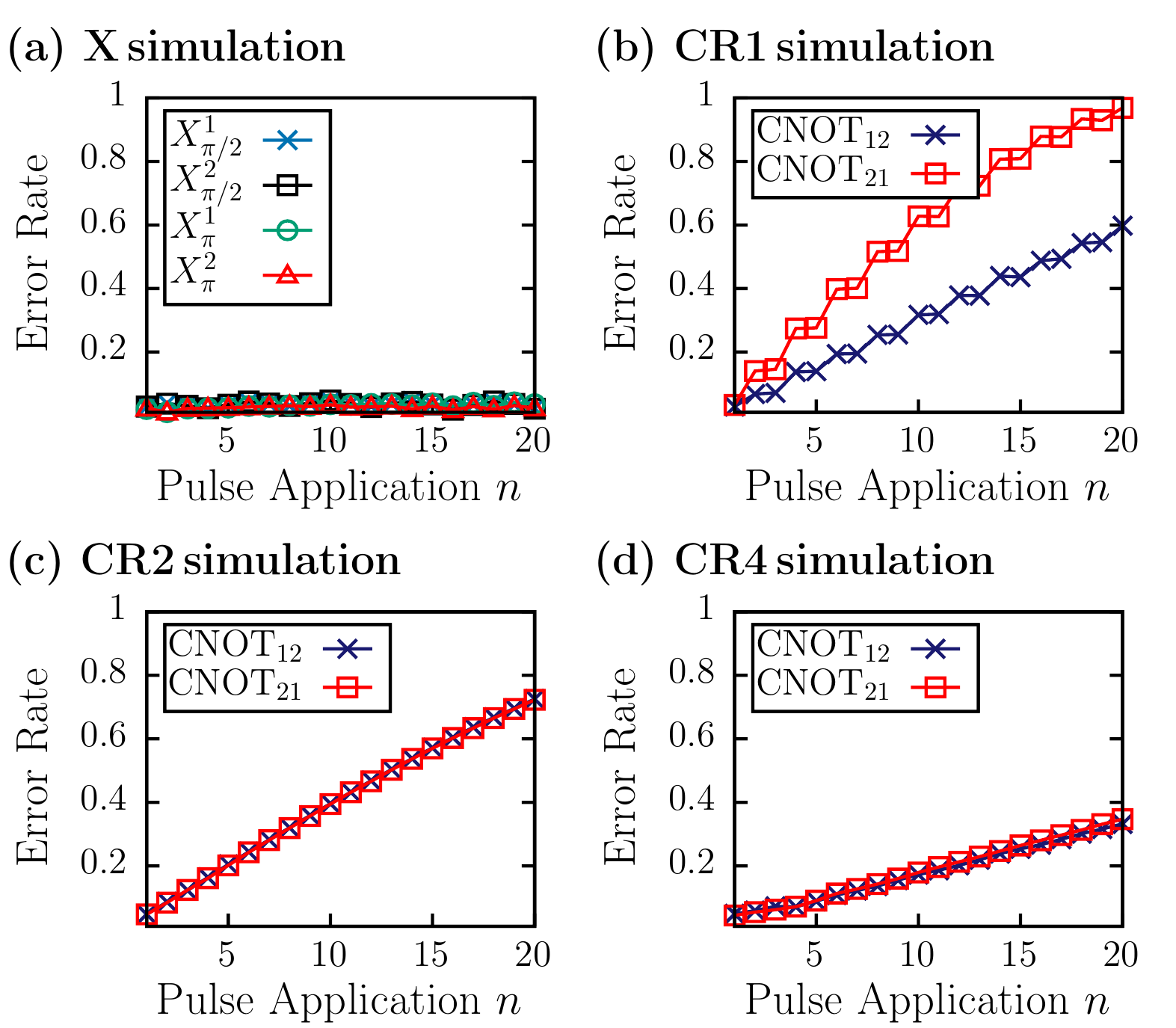}
  \caption{\label{fig:errorrates}(Color online) Evolution of the error rate given by \equref{eq:errorrate} after $n$ applications of a certain gate. (a) Single-qubit Gaussian derivative pulses defined by \equref{eq:singlequbitpulse}; (b)-(d) two-qubit CNOT gates based on the cross-resonance effect. While CR1 includes only one CR pulse on each of the transmons, CR2 and CR4 employ additional $X$ gates to echo out certain errors (see \secref{sec:model} and \appref{app:gatepulsedetails}).}
\end{figure}
\begin{table}
  \caption{\label{tab:qft}Comparison of the error rate $\eta_\Diamond$ for a single $\mathrm{CNOT}_{12}$ gate, twenty successive $\mathrm{CNOT}_{12}$ gates, and four successive QFT applications. A QFT contains five CNOT gates and two additional X pulses. The numbers reported are the error rates defined in \equref{eq:errorrate}, but the same relative trends are true for the average gate infidelity $1-F_{\text{avg}}$ given by \equref{eq:fidelity} and the unitarity $u$ given by \equref{eq:unitarity}.}
\begin{ruledtabular}
\begin{tabular}{@{}cccc@{}}
  Pulse & $\mathrm{CNOT}^1$ & $\mathrm{CNOT}^{20}$ & $\mathrm{QFT}^4$ \\
  \colrule\T
  \textbf{CR2} & 0.048 & 0.73 & 0.27 \\
  \textbf{CR4} & 0.049 & 0.33 & 0.32 \\
\end{tabular}
\end{ruledtabular}
\end{table}

For each gate primitive of our universal gate set, we study $n=1,\ldots,20$ repeated applications of the corresponding pulses on each of the four computational basis states. After each application of a pulse with total time $T$, we construct the full transformation matrix $M(nT)$ of the computational subspace as described in \secref{sec:pulseoptimization} and compare it with the ideal gate matrix $U^n$. Interestingly, we observed that the actual transformation $M(nT)$ is always closer to $U^n$ than the product $M(T)^n$, which means that the actual pulse performs better than the transformation $M(T)$ on the computational subspace suggests. However, this also means that the non-computational levels are more significant in the time evolution than a simple two-state description of a quantum computer can capture (see also \cite{Michielsen2017BenchmarkingQC}).

In \figref{fig:errorrates}, we plot the error rate $\eta_\Diamond$ given by \equref{eq:errorrate} to compare $M(nT)$ with $U^n$. We choose $\eta_\Diamond$ because this quantity is central for fault-tolerant quantum computation, and it also includes the statistical distance of the experimentally measurable output distribution \cite{Sanders2016ThresholdTheorem}. We observed the same qualitative behavior for the distance objective given by \equref{eq:distance} and the average gate infidelity $1-F_{\text{avg}}$ (data not shown).

The single-qubit pulses perform reasonably well. Although the error rates are always above $2\%$ (see \tabref{tab:metrics}), they stay approximately constant even after successive applications of the gates (see \figref{fig:errorrates}(a)). For the two-qubit gates, the error rate already starts growing after two applications of the CNOT gate (see \figref{fig:errorrates}(b),(c),(d)). This is most clearly visible for the CR1 gate, for which the error rate makes a jump after every second application. Note that the echoed CNOT gates CR2 and CR4 are found to work equally well if the control and the target qubit are exchanged. In experiments, usually only one type of CNOT is implemented because the CR interaction strength is weaker for the other type \cite{Takita2017faultTolerantStatePreparation,takita2016demonstration,ibmquantumexperience2016} (see also \figref{fig:CRamplitudescan} in \appref{app:gatepulsedetails}). The best performance is seen for the four-pulse echoed gate CR4, even though the gate metrics in \tabref{tab:metrics} do not suggest that. Note that the same discrepancy between worse metrics and better actual performance was also observed in recent quantum error-detection experiments on the IBMQX \cite{Takita2017faultTolerantStatePreparation}.

As an additional comparison between CR2 and CR4, we analyze both schemes in four applications of the quantum Fourier transform (QFT). The full circuit for $\mathrm{QFT}^4$ also involves 20 CNOT gates (along with eight additional X pulses, see \appref{app:programs}). Based on the error rates for 20 CNOT gates presented in \figref{fig:errorrates}, we might be led to believe that CR4 performs better in general. However, from the results presented in \tabref{tab:qft}, we see that this is not true. Hence, the error rate does not predict the behavior of a gate in actual applications. Note that the same is true for the average gate infidelity and the unitarity (data not shown).

It is worth mentioning that the gate with the worst fidelity and the worst unitarity ($\mathrm{CNOT}_{12}$ from the group CR1, see \tabref{tab:metrics}) is in fact the fastest and performs relatively well after repeated use, as demonstrated in \figref{fig:errorrates}(b). Similarly, the gate with the best fidelity ($\mathrm{CNOT}_{21}$ from the group CR1) performs worst. This means that, although the analyzed gate metrics can be used to study errors in one application of a gate, they do not characterize the performance of the gates when used in a quantum circuit (see also \cite{Michielsen2017BenchmarkingQC}).

\subsection{Comparison with the IBM Quantum Experience}
\begin{figure}
  \includegraphics[width=\linewidth]{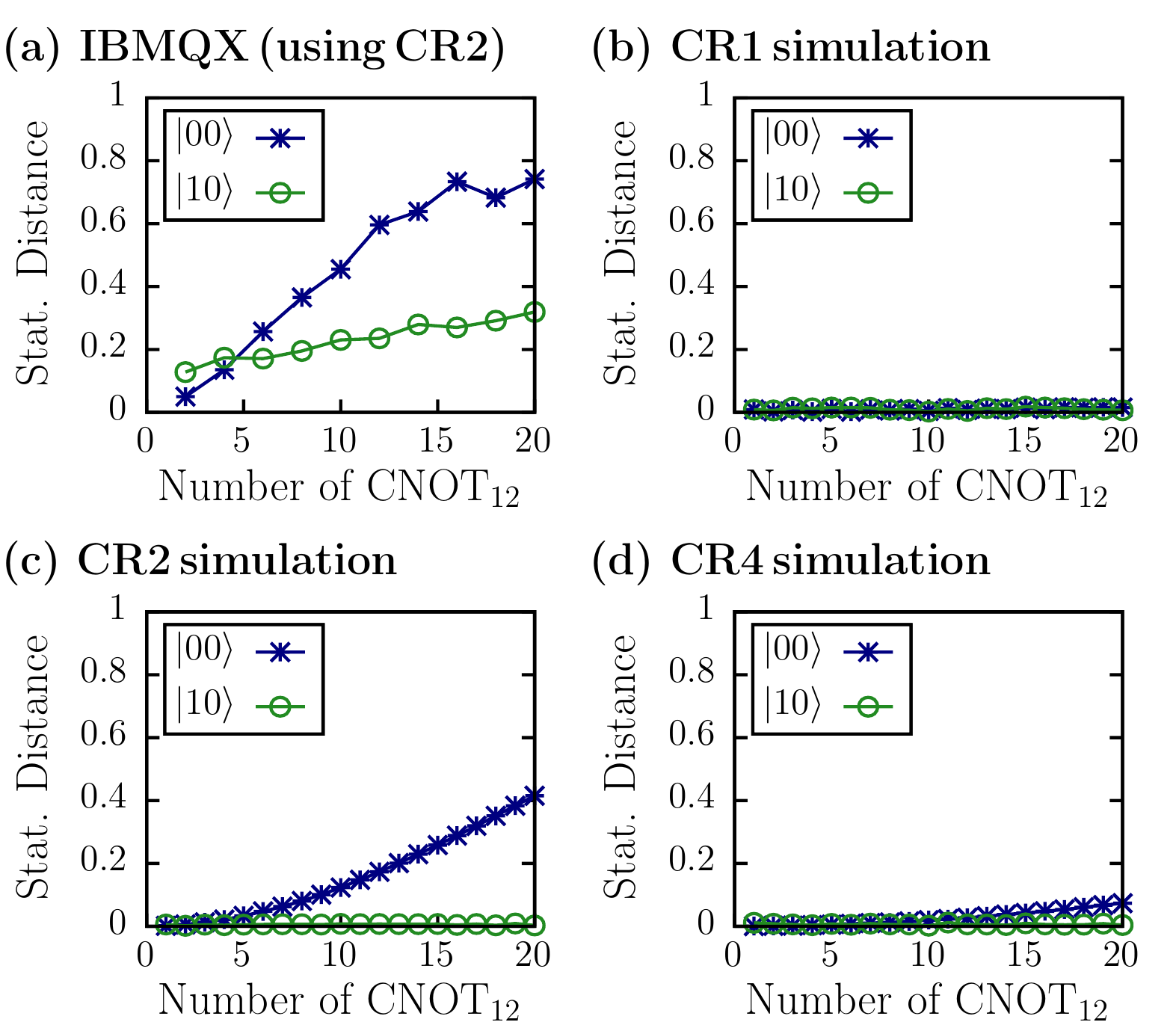}
  \caption{\label{fig:cnot}(Color online) Statistical distances between the ideal result and the measured distribution of states for the circuit $\mathrm{CNOT}_{12}^n\ket{\psi}$ with $\ket{\psi}=\ket{00}$ (stars) and $\ket{\psi}=\ket{10}$ (circles). (a) Experimental results on the IBMQX; (b)-(d) simulation results. Generically, the echoed CNOT versions show worse performance on state $\ket{00}$ than on state $\ket{10}$, both in the experiment and the simulation. Interestingly, this systematic error is not present in the one-pulse version CR1.}
\end{figure}
\begin{figure}
  \includegraphics[width=\linewidth]{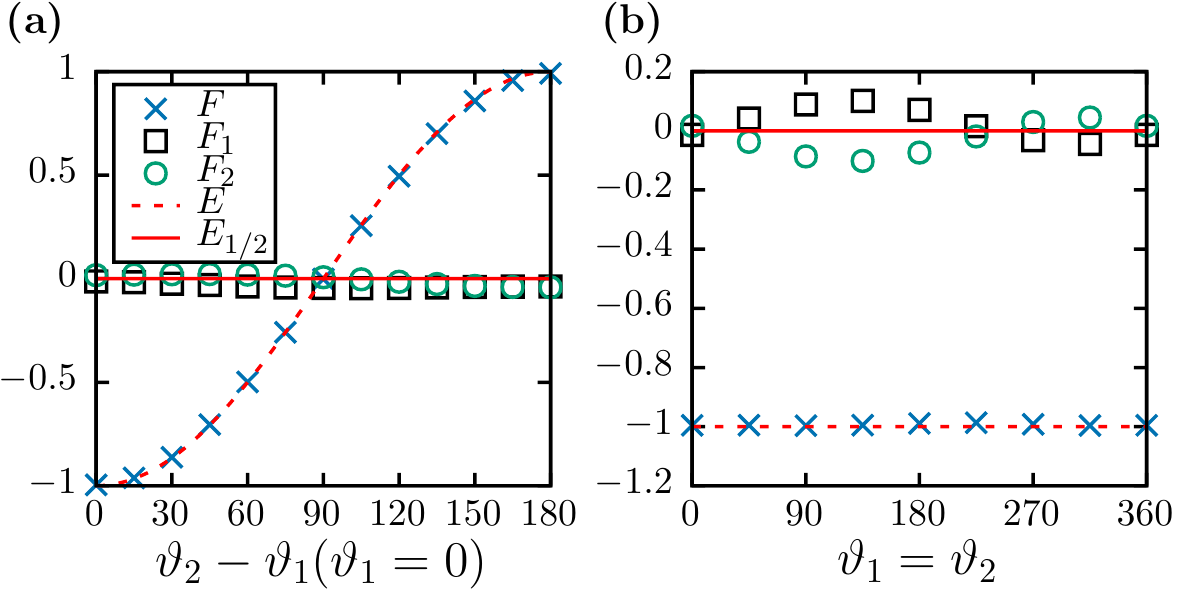}
  \caption{\label{fig:singlet}(Color online) Results for a set of quantum circuits creating and characterizing the singlet state $(\ket{01}-\ket{10})/\sqrt{2}$ as a function of the angles $\vartheta_1$ and $\vartheta_2$. $F_{1/2}(\vartheta_1,\vartheta_2)$ are single-qubit averages and $F(\vartheta_1,\vartheta_2)$ is a two-qubit correlation function. The corresponding theoretical expectations are given by $E_{1/2}(\vartheta_1,\vartheta_2)$ and $E(\vartheta_1,\vartheta_2)$. (a) $\vartheta_1=0$ fixed and $\vartheta_2$ variable; (b) $\vartheta_1=\vartheta_2$ variable. Apart from a small systematic deviation around $\vartheta_1=\vartheta_2=135$, the agreement is almost perfect.}
\end{figure}

As the simulation model is inspired by the quantum processor of the IBMQX, we perform two experiments to compare the simulation model with the physical hardware. In this way, the results of the simulated quantum processor can be directly compared to experimental results for a device using the same pulse schemes to implement quantum gates.

The first experiment again involves twenty successive CNOT gates, but this time we measure the statistical distance
\begin{align}
  D &= \frac{1}{2}\sum\limits_{i,j=0}^{1}|p_{ij} - \tilde{p}_{ij}|,
  \label{eq:statdistance}
\end{align}
between the ideal outcome distribution $p_{ij}$ (i.e., the probability to measure the state $\ket{ij}$) and the experimentally measured relative frequencies $\tilde{p}_{ij}$. The experiment on the IBMQX was performed with $8192$ shots on August 17, 2017 using Q3 as the control and Q4 as the target qubit. The results are presented in \figref{fig:cnot}.

The simulated CR2 gate gives the best qualitative agreement between experiment and simulation. This makes sense because the CR2 pulse scheme shown in \figref{fig:CNOTpulses} is also used for the IBMQX \cite{ibmquantumexperience2016}. Most remarkable is the fact that the performance for the initial state $\ket{00}$ is much worse than for $\ket{10}$ (see also Table 3 in \cite{Michielsen2017BenchmarkingQC}). As this also shows up in the ideal simulation, it points to a systematic error in the implementation of the CNOT gate. Note that this error is only present in the echoed gates, and not in the proposed one-pulse gate CR1. The remaining difference between the CR2 simulation and the experiment may be due to decoherence by the environment; we leave the study of including a heat bath in the simulation to future research.

As a second experiment, we repeat the two-qubit entanglement experiments proposed as part of a benchmark for gate-based quantum computers \cite{Michielsen2017BenchmarkingQC}. The quantum circuits first create the maximally entangled singlet state $(\ket{01}-\ket{10})/\sqrt{2}$ and then apply a set of rotations dependent on the angles $\vartheta_1$ and $\vartheta_2$ to analyze the constructed state (see \appref{app:programs} for the circuit). We select the two-pulse echoed CR2 gate for the CNOT operation, as also done for the IBMQX. We parse programs formulated in a quantum assembly language similar to the one used by the IBMQX \cite{Cross2017openqasm2} to run the quantum circuits. The results are shown in \figref{fig:singlet}.

Although the gates used in the simulation, which are in some sense ideal versions of the gates used in experiments, do not reach perfect fidelities or error rates (see \tabref{tab:metrics}), they still yield almost perfect results for the entanglement experiments. The results are much closer to those expected for a singlet state than the corresponding experimental results on the IBMQX \cite{Michielsen2017BenchmarkingQC}, even though the reported fidelities of the latter are the same or even better. This can have three reasons: (i) the procedure of measuring the fidelities (i.e., randomized benchmarking) produces numbers that overestimate the gate performance (cf. \cite{proctor2017RandomizedBenchmarkingCriticism,Wallman2017RandomizedBenchmarkingCriticism,Sanders2016ThresholdTheorem}), implying that the actual gate implementations are worse; (ii) the actual gates are good but the discrepancy is due to another process (such as the measurement) that is not yet included in the simulation model; or (iii) other unknown factors not included in the quantum-theoretical description of the experiments play a destructive role.

\section{Discussion}\label{sec:discussion}
We have implemented algorithms to solve the TDSE for a quantum-mechanical model of superconducting transmon qubits coupled by transmission-line resonators. The architecture of the publicly accessible quantum computer by IBM is of this type \cite{ibmquantumexperience2016}. Great care has been taken to make no approximations to the Hamiltonian obtained from the circuit quantization \cite{koch2007transmon}.

The tested quantum gates are realized by applying Gaussian microwave pulses to the system, with pulse parameters determined by an optimization routine. Hence, we are confident that they represent idealized versions of the pulses used in recent experiments for this architecture. This is confirmed by almost perfect results for the entanglement experiment \cite{Michielsen2017BenchmarkingQC}. Thus, our simulation can be seen as an ideal version of the experiment. Still, the fact that all of our apparently ideal gates show diamond error rates above 2\%, suggests that the goal of building a universal, fault-tolerant quantum computer still remains a difficult, ongoing challenge.

We have found that gate metrics such as the average gate fidelity \cite{nielsen2002gatefidelity}, the diamond distance \cite{kitaev1997diamondnorm,Sanders2016ThresholdTheorem}, and the unitarity \cite{Wallman2015unitarity} each provide insights into the errors of the implemented gate pulses. Specifically, while the time evolution of the total system is inherently unitary and the errors are systematic, they appear as incoherent non-Pauli errors on the computational subspace. Conceptually, these errors originate from entanglement between the computational states and the non-computational states in the transmons and the resonator. Such errors cause most of the mismatch between the ideal gates and the implemented pulses (see also \cite{Kueng2016ComparingExperimentsToThreshold,ghosh2013understandingtheeffectsofleakage,Sanders2016ThresholdTheorem,Wallman2015unitarity,Magesan2013ModelingQuantumNoise,Puzzuoli2014tractablesimulation}).

However, the information obtained from the gate metrics is not enough to assess the error induced by repeatedly using the gate in quantum algorithms. To be precise, a gate showing close-to-ideal performance with respect to the studied gate metrics can still perform worse than an initially less ideal gate after multiple applications. In particular, the entangling two-qubit gates were found to lose performance over repeated applications. Especially the two-pulse echoed CR2 gate exhibited systematic errors that could also be observed in the IBMQX. In comparison, the longer CR4 gate seemed to perform better in spite of worse fidelity, as also seen in recent experiments \cite{takita2016demonstration,Takita2017faultTolerantStatePreparation}. When used in a QFT algorithm, however, this observation was reversed again. An extreme case was given by the one-pulse CNOT gate CR1, which for $\mathrm{CNOT}_{21}$ gave the best fidelity but the worst performance. In contrast, $\mathrm{CNOT}_{12}$ showed the worst fidelity and the worst unitarity but a reasonably good performance, without suffering from the systematic error present in CR2 and CR4. Hence, the gate metrics under investigation do not provide reliable information of how well and how often a certain gate may be used in an algorithm (see also the conclusion in \cite{Darmawan2016MetricsForQEC}). As this information is essential for potential users of gate-based quantum computers, it should be included in the specification sheet of the physical device.

Future work will go into scaling up the simulation to model experiments with more qubits and additional coupling schemes, in accordance with the goal pursued in experiments. This then enables a detailed simulation of quantum error-correcting codes under realistic conditions for various architectures. In addition to that, we plan to simulate the measurement process in detail.

\begin{acknowledgments}
We are grateful to the IBM Quantum Experience project team for sharing technical details with us.
This work does not reflect the views or opinions of IBM or any of its employees.
\end{acknowledgments}

\appendix
\section{Description of the algorithm}\label{app:algorithm}
The algorithm that we employ to solve the TDSE given by \equref{eq:TDSE} is a Suzuki-Trotter product-formula algorithm constructed from the Hamiltonian by using the general framework presented in \cite{deraedt1987productformula}. The algorithm is explicit, inherently unitary, and unconditionally stable by construction. Among others, the framework has been used to devise algorithms for NMR systems for quantum computation \cite{deraedt2004computational}, and it also forms the basis of the massively parallel quantum computer simulator \cite{DeRaedt2007MassivelyParallel} that can nowadays simulate systems with up to 45 qubits \cite{Ito2017MassivelyParallel}.

The model Hamiltonian $H$ given by \equref{eq:modelhamiltonian} needs to be expressed in an appropriate basis to derive the algorithm. In this work, we choose the charge basis $\{\ket{n_1n_2}:n_i\in\bb Z\}$ (i.e. the joint eigenbasis of the number operators $\hat{n}_1$ and $\hat{n}_2$) for the qubits and the Fock basis $\{\ket{k}:k\in\bb N_0\}$ for the resonator. This basis has the nice property that $H$ can be generically expressed as a sum of tensor products of tridiagonal matrices.

At the heart of the algorithm lies a decomposition of the total unitary time-evolution operator
\begin{align}
  U_{\text{total}}(t',t)&=\mc T \exp\!\left(-i\int_{t}^{t'} \da\tau H(\tau)\right),
  \label{eq:fulltimeevolutionoperator}
\end{align}
where $\mc T$ is the time-ordering symbol. This expression is first discretized in time steps $\tau$, i.e., we consider the propagator $U_{t+\tau,t}=\exp(-i\tau H(t+\tau/2))$. Note that $\tau$ needs to be chosen small enough with respect to the energy scales and the other relevant time scales of $H(t)$ such that the exact mathematical solution of the TDSE is obtained up to some fixed numerical precision. Subsequently, the exponential of $H(t+\tau/2)$ is decomposed using the Lie-Trotter-Suzuki product-formula \cite{Suzuki1977}. This is done by partitioning the tridiagonal matrices into even and odd sums of $2\times2$ block-diagonal matrices such that each matrix exponential can be evaluated analytically (cf.~\cite{deraedt1987productformula,deraedt2004computational}). With these, we iteratively update the state vector $\ket{\Psi(t)}=\sum a_{kn_1n_2}(t)\ket{k}\!\ket{n_1}\!\ket{n_2}$ using the second-order expression of the framework. Finally, the solution is transformed to the transmon basis $\{\ket{m_1m_2}:m_i\in\bb N_0\}$ (see \equref{eq:TDSEsolution}) by computing $a_{km_1m_2}(t) = \sum_{n_1n_2} (B_{n_1m_1}^1)^* (B_{n_2m_2}^2)^* a_{kn_1n_2}(t)$, where the $B_{n_im_i}^i$ are defined by $\ket{m_i}=\sum_{n_i} B_{n_im_i}^i \ket{n_i}$ and obtained from the eigenvectors of $H_{CPB}$ given by \equref{eq:modelhamiltonianCPB} for $n_{gi}(t)=0$.

In practice, we set the time step to solve the TDSE to $\tau=\SI{0.1}{ps}$ and the number of states included in the product basis to $n_i=-8,\ldots,8$ and $k=0,\ldots,3$. We stress that no further approximation needs to be made to obtain the solution of the TDSE.

The software is written in C++ and the implementation of the algorithms has been validated by comparison with exact diagonalization for smaller Hilbert spaces. Furthermore, we have checked that the results are qualitatively independent of small variations in the time step $\tau$, the number of charge and photon states included in the basis, and the particular device parameters given in \tabref{tab:device}.

\section{Details about the gate pulses}\label{app:gatepulsedetails}
The quantum gate set that we physically implement reads
\begin{align}
  \mc S = \{X^1_{\pi/2},X^2_{\pi/2},X^1_{\pi},X^2_{\pi},\mathrm{CNOT}_{12},\mathrm{CNOT}_{21}\},
  \label{eq:gateset}
\end{align}
where $X^j_{\varphi}=\exp(-i\varphi \sigma_j^x/2)$ is a rotation of qubit $j$ about the $x$ axis by an angle of $\varphi$, and $\mathrm{CNOT}_{ij}$ is defined by negating the target qubit $j$ if the control qubit $i$ is in the state $\ket{1}$ and doing nothing if it is in the state $\ket{0}$. We additionally support the VZ gates $Z^j_{\varphi}=\exp(-i\varphi\sigma_j^z/2)$ for arbitrary angles $\varphi$ to make the gate set $\mc S$ universal for quantum computation \cite{NielsenChuang}. By analogy with experiments, a VZ gate does not correspond to a separate pulse, but it changes the phases of all the following pulses (see \cite{McKay2016VZgate}). For this purpose, we keep track of two offset phases $\phi_1$ and $\phi_2$ during the evolution, and the phase $\gamma$ in \equref{eq:genericpulses} of every subsequent pulse oscillating at $\bar\omega_1$ ($\bar\omega_2$) is shifted by $-\phi_1$ ($-\phi_2$).

We also employ these zero-duration VZ gates to correct phase errors in the gate sequence resulting from phase shifts due to other non-computational levels or off-resonant driving \cite{McKay2016VZgate}. In particular, this means that each gate is followed by a local Z rotation of the form $Z_{\varphi_1}\otimes Z_{\varphi_2}$ which essentially only results in an update of the tracked phases. The phases $\varphi_1$ and $\varphi_2$ are found by an additional optimization step using the objective function given by \equref{eq:distance}, but this time replacing the result $M$ of the first optimization by $(Z_{\varphi_1}\otimes Z_{\varphi_2})M$. While this does not change a single gate before the measurement, and the optimized correction phases $\varphi_1$ and $\varphi_2$ are close to $0$, we have observed that it considerably improves the performance of the gates after repeated applications because it mitigates the accumulation of phase errors.

\subsection{Single-qubit gates}
\begin{table}
  \caption{\label{tab:singlequbitpulses}Parameters defining the single-qubit pulses as obtained by the pulse optimization with the initial values taken from the theory of transmon qubit control \cite{gambetta2013controlIFF}.}
\begin{ruledtabular}
\begin{tabular}{@{}lrrrr@{}}
  \multicolumn{1}{c}{Pulse} & \multicolumn{1}{c}{$\Omega_0$} & \multicolumn{1}{c}{$\beta$ in ns} & \multicolumn{1}{c}{$\varphi_1$} & \multicolumn{1}{c}{$\varphi_2$}\\
  \colrule\T
  $\func{GD}^1_{\pi/2}$ & 0.00222 & 0.231 & $-$0.00202  & \hphantom{$-$}0.00328 \\
  $\func{GD}^2_{\pi/2}$ & 0.00227 & 0.289 & $-$0.00013 & $-$0.00159 \\
  $\func{GD}^1_{\pi}$   & 0.00444 & 0.219 & $-$0.00354 & \hphantom{$-$}0.00283 \\
  $\func{GD}^2_{\pi}$   & 0.00454 & 0.224 & $-$0.00026 & $-$0.00339 \\
\end{tabular}
\end{ruledtabular}
\end{table}

The general pulse for a single qubit gate is given by \equref{eq:singlequbitpulse} and depends on the parameters $(\Omega_0,\beta,\gamma)$. The amplitude $\Omega_0$ is directly proportional to the implemented angle of rotation $\vartheta\in\{\pi/2,\pi\}$ and can be obtained from the relation $\vartheta=b_i\int_0^T\Omega_G(t)\ds t$ where $b_i=2E_{Ci}(E_{Ji}/8E_{Ci})^{1/4}$ \cite{gambetta2013controlIFF}. The DRAG coefficient $\beta$ is initially set to $-1/2\alpha_i$ (see \cite{gambetta2010dragtheory}), where the anharmonicity $\alpha_i$ is given in \tabref{tab:device}. These two parameters are refined in the optimization procedure as described in \secref{sec:pulseoptimization}.

In the following, we denote the resulting single-qubit pulses on qubit $i$ by $\func{GD}^i_{\pi/2}(\gamma)$ and $\func{GD}^i_{\pi}(\gamma)$. The corresponding pulse parameters along with the above-mentioned VZ phase corrections $\varphi_1$ and $\varphi_2$ are given in \tabref{tab:singlequbitpulses}. The only parameter left in these pulses is the phase $\gamma$. This phase is used to implement VZ gates according to the scheme \cite{McKay2016VZgate}
\begin{align}
  \label{eq:VZgatephaseshift}
  \func{GD}_\vartheta^i(\gamma)\,Z_{\varphi}\ket{\psi} = Z_{\varphi}\,\func{GD}_{\vartheta}^i(\gamma-\varphi)\ket{\psi}.
\end{align}

\subsection{Two-qubit gates}
\begin{table}
  \caption{\label{tab:twoqubitpulses}Parameters defining the two-qubit pulses, resulting from the pulse optimization procedure. The parameters $\varphi_{\text{CR}}$, $\Omega_{\text{Cancel}}$, and $\varphi_{\text{Cancel}}$ are only needed for the CR1 scheme.}
\begin{ruledtabular}
\begin{tabular}{@{}lccccccc@{}}
  \multicolumn{1}{c}{Pulse} & \multicolumn{1}{c}{$T_{\text{CR}}$ in ns} & \multicolumn{1}{c}{$\Omega_{\text{CR}}$} & \multicolumn{1}{c}{$\varphi_{\text{CR}}$} & \multicolumn{1}{c}{$\Omega_{\text{Cancel}}$} & \multicolumn{1}{c}{$\varphi_{\text{Cancel}}$} & \multicolumn{1}{c}{$\varphi_1$} & \multicolumn{1}{c}{$\varphi_2$}\\
  \colrule\T
  $\func{GF}^1_{\func{CR1}}$ & \hphantom{0}41.86 & 0.079 & \hphantom{$-$}0.54 & \hphantom{$-$}0.0062 & \hphantom{$-$}0.00 &            $-$2.10 & \hphantom{$-$}0.04 \\
  $\func{GF}^2_{\func{CR1}}$ & 128.19            & 0.094 &            $-$2.89 &            $-$0.0016 & \hphantom{$-$}1.72 & \hphantom{$-$}3.25 & \hphantom{$-$}1.40 \\
  $\func{GF}^1_{\func{CR2}}$ & 102.97            & 0.011 & $-$ & $-$ & $-$ & \hphantom{$-$}0.00 & \hphantom{$-$}0.00 \\
  $\func{GF}^2_{\func{CR2}}$ & \hphantom{0}71.56 & 0.071 & $-$ & $-$ & $-$ & \hphantom{$-$}0.00 & \hphantom{$-$}0.00 \\
  $\func{GF}^1_{\func{CR4}}$ & \hphantom{0}50.24 & 0.010 & $-$ & $-$ & $-$ & \hphantom{$-$}0.00 &            $-$0.01 \\
  $\func{GF}^2_{\func{CR4}}$ & \hphantom{0}30.16 & 0.069 & $-$ & $-$ & $-$ &            $-$0.01 & \hphantom{$-$}0.00 \\
\end{tabular}
\end{ruledtabular}
\end{table}
\begin{figure}
  \includegraphics[width=\linewidth]{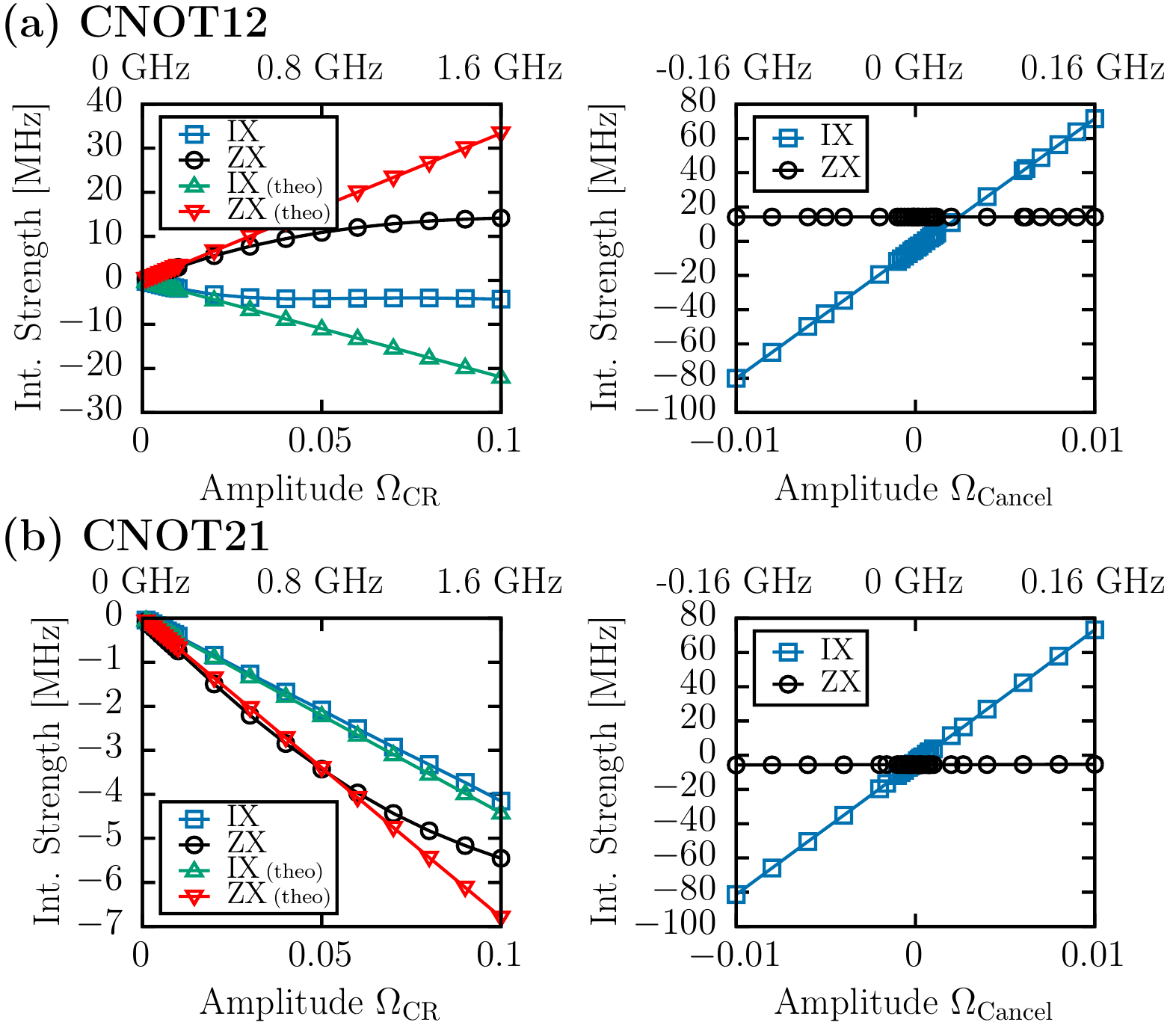}
  \caption{\label{fig:CRamplitudescan}(Color online) Scan of the CR drive amplitudes $\Omega_{\text{CR}}$ on the control qubit and $\Omega_{\text{Cancel}}$ on the target qubit. The dimensionless amplitudes can be converted to the strength of the drive by multiplying them with $b_i=2E_{Ci}(E_{Ji}/8E_{Ci})^{1/4}$ (shown on top of the plots). The IX and ZX interaction strengths are inferred by measuring the oscillations of the target qubit conditional on the control qubit being in state $\ket{0}$ and state $\ket{1}$ \cite{sheldon2016procedure}. The linear theory predictions can be derived perturbatively \cite{gambetta2013controlIFF} and are only valid for weak drivings. Note that the additional drive on the target qubit (the two figures on the right, shown for $\Omega_{\text{CR}}=0.1$ fixed) linearly displaces IX only. Thus, it can either be tuned to single out ZX or to generate the CNOT gate directly up to local Z rotations.}
\end{figure}
The central pulse in two-qubit gates for the present architecture is the cross-resonance (CR) pulse, depicted as a flat-topped Gaussian in \figref{fig:CNOTpulses}. It always oscillates at the frequency of the target qubit $\bar\omega_T$ and it is defined by its amplitude $\Omega_{\text{CR}}$ and the time $T_{\text{CR}}$ of the flat top (thus the time of the CR drive including rise and fall is $T_{\text{CR}}+\SI{30}{ns}$). The CR1 scheme additionally includes the amplitude of the target drive $\Omega_{\text{Cancel}}$ and two phases $\varphi_{\text{CR}}$ and $\varphi_{\text{Cancel}}$, inspired by the observations in \cite{sheldon2016procedure}.

Although there are theoretical predictions based on perturbation theory for the specific choice of parameters \cite{gambetta2013controlIFF,sheldon2016procedure}, they need to be fine-tuned to the specific set of qubits. We do this by scanning the amplitudes for a CR drive and obtaining the CR interaction strengths from the conditional rotation of the target qubit, as done in \cite{sheldon2016procedure}. Such a scan is shown in \figref{fig:CRamplitudescan}. As the initial goal of CR gates was to single out a ZX interaction \cite{Corcoles2013processverification}, the CR2 and CR4 gates use an echo scheme to echo out the IX interaction. The one-pulse gate CR1, in contrast, uses the additional drive on the target qubit to shift IX such that the implemented transformation is $\exp(-i\pi(3\sigma_T^x+\sigma_C^z\sigma_T^x)/4)$, which is equal to a CNOT gate up to local Z rotations. The correct time $T_{\text{CR}}$ for each pulse is obtained from a separate scan.

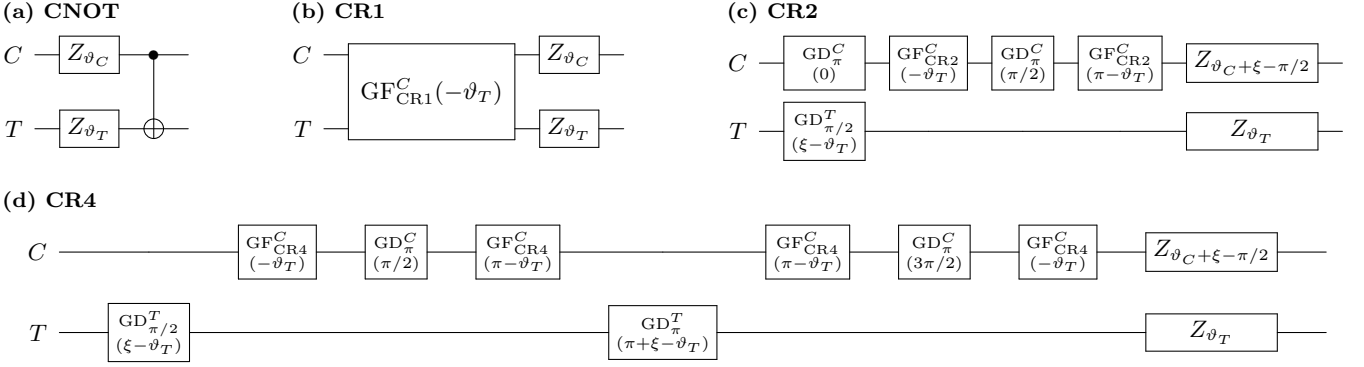
\begin{figure*}
  \captionsetup[subfigure]{position=top,labelfont=bf,textfont=normalfont,singlelinecheck=off,justification=raggedright}
  \subfloat[\label{sfig:CNOT}\bfseries CNOT]{
    $
      \Qcircuit @C=1em @R=1.5em {
       & \lstick{C} & \gate{Z_{\vartheta_C}} & \ctrl{1} & \qw \\
       & \lstick{T} & \gate{Z_{\vartheta_T}} & \targ & \qw \\
      }
    $
  }\hfill
  \subfloat[\label{sfig:CR1}\bfseries CR1]{
    $
      \Qcircuit @C=1em @R=1.5em {
	& \lstick{C} & \multigate{1}{\text{GF}_{\text{CR1}}^C(-\vartheta_T)} & \gate{Z_{\vartheta_C}} & \qw \\
        & \lstick{T} & \ghost{\text{GF}_{\text{CR1}}^C(-\vartheta_T)} & \gate{Z_{\vartheta_T}} & \qw \\
      }
    $
  }\hfill
  \subfloat[\label{sfig:CR2}\bfseries CR2]{
    $
      \Qcircuit @C=1em @R=.5em {
	& \lstick{C} & \gate{\makesamesize{\substack{\text{GD}_{\pi/2}^T\\(\xi-\vartheta_T)}}{\substack{\text{GD}_{\pi}^C\\(0)}}} & \gate{\substack{\text{GF}_{\text{CR2}}^C\\(-\vartheta_T)}} & \gate{\substack{\text{GD}_{\pi}^C\\(\pi/2)}} & \gate{\substack{\text{GF}_{\text{CR2}}^C\\(\pi-\vartheta_T)}} & \gate{Z_{\vartheta_C+\xi-\pi/2}} & \qw \\
	& \lstick{T} & \gate{\substack{\text{GD}_{\pi/2}^T\\(\xi-\vartheta_T)}} & \qw & \qw & \qw & \gate{\makesamesize{Z_{\vartheta_C+\xi-\pi/2}}{Z_{\vartheta_T}}} & \qw \\
      }
    $
  }\\
  \subfloat[\label{sfig:CR4}\bfseries CR4]{
    $
      \Qcircuit @C=2em @R=1em {
	& \lstick{C} & \qw & \gate{\substack{\text{GF}_{\text{CR4}}^C\\(-\vartheta_T)}} & \gate{\substack{\text{GD}_{\pi}^C\\(\pi/2)}} & \gate{\substack{\text{GF}_{\text{CR4}}^C\\(\pi-\vartheta_T)}} & \qw & \gate{\substack{\text{GF}_{\text{CR4}}^C\\(\pi-\vartheta_T)}} & \gate{\substack{\text{GD}_{\pi}^C\\(3\pi/2)}} & \gate{\substack{\text{GF}_{\text{CR4}}^C\\(-\vartheta_T)}} & \gate{Z_{\vartheta_C+\xi-\pi/2}} & \qw \\
	& \lstick{T} & \gate{\substack{\text{GD}_{\pi/2}^T\\(\xi-\vartheta_T)}} & \qw & \qw & \qw & \gate{\substack{\text{GD}_{\pi}^T\\(\pi+\xi-\vartheta_T)}} & \qw & \qw & \qw & \gate{\makesamesize{Z_{\vartheta_C+\xi-\pi/2}}{Z_{\vartheta_T}}} & \qw \\
      }
    $
  }\hfill\null
  \caption{\label{fig:twoqubitcircuits}Specifications of the pulse sequences in \figref{fig:CNOTpulses} to implement a generic CNOT gate with VZ phases. The elementary Gaussian pulse GD is defined in \tabref{tab:singlequbitpulses}, and GF is defined in \tabref{tab:twoqubitpulses}. C (T) is the control (target) qubit. \protect\subref{sfig:CNOT} Generic CNOT gate with the preceding VZ phases that all pulses need to be capable of shifting through; \protect\subref{sfig:CR1} one-pulse CR1 gate which includes a flat-topped Gaussian pulse on the control and the target qubit simultaneously; \protect\subref{sfig:CR2} two-pulse echoed CR2 gate; \protect\subref{sfig:CR4} four-pulse echoed CR4 gate. In the CR2 and the CR4 scheme, the additional phase shift $\xi$ is zero if $\bar\omega_C>\bar\omega_T$ and $\pi$ otherwise to handle the case when ZX is negative (see \figref{fig:CRamplitudescan}).}
\end{figure*}

The final pulse parameters are then found in the pulse optimization procedure (see \secref{sec:pulseoptimization}). By analogy with the single-qubit pulses, we denote the flat-topped CR drivings on qubit $i$ by $\func{GF}_{CR*}^i(\gamma)$. The corresponding parameters and the VZ phase corrections are given in \tabref{tab:twoqubitpulses}. Again, $\gamma$ is the only variable parameter, and it can be used to implement VZ gates in the same way as in \equref{eq:VZgatephaseshift}. Note that, as the CNOT gate commutes with Z gates on the control qubit, only phase shifts of the target qubit affect $\gamma$ \cite{McKay2016VZgate}. The full specifications including the scheme to implement VZ gates are given in \figref{fig:twoqubitcircuits}.

\section{Circuits for the quantum programs}\label{app:programs}

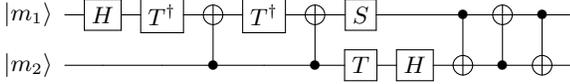
\begin{figure}
  \[
    \Qcircuit @C=.8em @R=.7em {
      &\lstick{\ket{m_1}}&\gate{H}&\gate{T^\dag}&\targ    &\gate{T^\dag}&\targ    &\gate{S}&\qw     &\ctrl{1}&\targ    &\ctrl{1}&\qw\\
      &\lstick{\ket{m_2}}&\qw     &\qw          &\ctrl{-1}&\qw          &\ctrl{-1}&\gate{T}&\gate{H}&\targ   &\ctrl{-1}&\targ   &\qw\\
    }
  \]
  \caption{\label{fig:circuitQFT}Circuit for the two-qubit QFT.}
\end{figure}

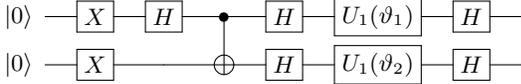
\begin{figure}
  \[
    \Qcircuit @C=1.3em @R=.4em {
      &\lstick{\ket{0}}&\gate{X}&\gate{H}&\ctrl{1}&\gate{H}&\gate{U_1(\vartheta_1)}&\gate{H}&\qw\\
      &\lstick{\ket{0}}&\gate{X}&\qw     &\targ   &\gate{H}&\gate{U_1(\vartheta_2)}&\gate{H}&\qw\\
    }
  \]
  \caption{\label{fig:circuitSinglet}Circuit for experiments on the singlet state.}
\end{figure}

In the following, we show the quantum circuits for the QFT algorithm and the entanglement experiments from \secref{sec:results}. The gates contained in the circuits map to the pulses defined in \appref{app:gatepulsedetails} in the same way as for the IBMQX \cite{Cross2017openqasm2}. In particular, we have $H=Z_{\pi/2}X_{\pi/2}Z_{\pi/2}$, $S=Z_{\pi/2}$, $T=Z_{\pi/4}$, $T^\dagger=Z_{-\pi/4}$, and $U_1(\vartheta)=Z_{\vartheta}$ (up to global phases).

The two-qubit QFT in principle contains two Hadamard gates $H$, one controlled-$S$ gate, and one SWAP gate \cite{NielsenChuang}. Rewriting this in terms of the gates supported by our system leads to the circuit given in \figref{fig:circuitQFT}. As only the $H$ gate and the CNOT gate result in actual hardware pulses, this circuit involves two X pulses and five CNOT pulses in total.

The circuit to analyze the singlet state as a function of the angles $\vartheta_1$ and $\vartheta_2$ is taken directly from \cite{Michielsen2017BenchmarkingQC} and is given in \figref{fig:circuitSinglet}.

\FloatBarrier
\bibliographystyle{apsrev4-1}
\bibliography{bibliography}

\end{document}